%% file: ASPLOS_main.tex
\documentclass[sigplan,nonacm]{acmart}

\AtBeginDocument{%
  }

\usepackage{listings}
\lstdefinelanguage{yaml}{
  keywords={true,false,null,y,n},
  keywordstyle=\color{blue}\bfseries,
  basicstyle=\ttfamily\small,
  commentstyle=\color{gray}\itshape,
  stringstyle=\color{red},
  morecomment=[l]{\#},
  morestring=[b]',
  morestring=[b]",
  showstringspaces=false
}
\lstdefinestyle{mystyle}{
  basicstyle=\ttfamily\footnotesize,
  frame=single,
  numbers=left,
  numberstyle=\tiny\color{gray},
  breaklines=true,
  tabsize=2,
  keywordstyle=\color{blue!60!black},
  commentstyle=\color{green!40!black},
  stringstyle=\color{orange!60!black},
  backgroundcolor=\color{gray!5}
}
\usepackage{subcaption}
\usepackage{algorithm}
\usepackage{enumitem}
\usepackage{booktabs}
\usepackage{tikz}
\usepackage{float}
\usetikzlibrary{arrows.meta, positioning}
\usepackage{pifont}
\newcommand{\cmark}{\ding{51}}
\newcommand{\xmark}{\ding{55}}
\usepackage{xspace}
\usepackage{tikz}
\usetikzlibrary{arrows.meta,positioning,fit,backgrounds}
\usepackage{pifont}
\usepackage{graphicx, xcolor}
\usepackage{makecell}
\usepackage{multirow}
\usepackage{dblfloatfix}

\usepackage{enumitem}
\setitemize{noitemsep,topsep=0pt,parsep=0pt,partopsep=0pt}

\definecolor{dong}{RGB}{0,0,200}
\definecolor{check}{RGB}{0,0,0}
\definecolor{jie}{RGB}{255,140,0}
\definecolor{charlie}{RGB}{0,201,87}
\definecolor{checked}{RGB}{0,0,0}
\definecolor{draft}{RGB}{102,204,0}
\definecolor{codeorange}{RGB}{255,140,0}
\newcommand{\shuangyan}[1]{\textcolor{codeorange}{#1}}
\newcommand{\name}{BloomBee\xspace}

\begin{document}
\title{Distributed Generative Inference of LLM at Internet Scales with Multi-Dimensional Communication Optimization}



\author{Jiu Chen}
\authornote{Equal contribution}
\affiliation{%
  \institution{University of California Merced}
  \country{USA}
}

\author{Shuangyan Yang}
\authornotemark[1]
\affiliation{%
  \institution{University of California Merced}
  \country{USA}
}

\author{Xu Xiong}
\authornotemark[1]
\affiliation{%
  \institution{University of California Merced}
  \country{USA}
}

\author{Hexiao Duan}
\affiliation{%
  \institution{University of California Merced}
  \country{USA}
}

\author{Xinran Zhang}
\affiliation{%
  \institution{University of California Berkeley}
  \country{USA}
}

\author{Jie Ren}
\affiliation{%
  \institution{College of William \& Mary}
  \country{USA}
}

\author{Dong Li}
\affiliation{%
    \institution{University of California Merced}
    \country{}
}
\affiliation{%
    \institution{Yotta Labs}
    \country{USA}
}

\renewcommand{\shortauthors}{Trovato et al.}





\input{text/abstract}

\maketitle

\input{text/intro}
\input{text/background}

\input{text/motivation}

\input{text/overview}
\input{text/design}

\input{text/implementation}
\input{text/evaluation}

\input{text/related_work}
\input{text/conclusion}

\clearpage
\bibliographystyle{ACM-Reference-Format}
\bibliography{bib/charlie,bib/li,bib/yang}

\appendix
\input{text/appendix}

\end{document}

%% file: text/abstract.tex
\begin{abstract}
    Decentralized LLM inference distributes computation among heterogeneous nodes across the internet, offering a performant and cost-efficient solution, alternative to traditional centralized inference. However, the low cross-node network bandwidth makes communication the primary bottleneck. In this paper, we introduce \name, an internet-scale distributed LLM inference framework. \name integrates LLM-layer assignment, micro-batching and tensor offloading to optimize communication from multiple dimensions. Additionally, \name formulates the coordination of these techniques as an optimization problem and solves it using dynamic programming. \name also customizes lossless compression and speculative decoding according to low-bandwidth network settings to reduce communication overhead. \textcolor{checked}{We evaluate \name across a spectrum of network environments and show that it improves service throughput by up to 1.76$\times$. It also reduces average latency by up to 43.20\% compared to state-of-the-art decentralized LLM inference systems. \name is open-sourced\footnote{https://github.com/ai-decentralized/BloomBee}.}
\end{abstract}

%% file: text/intro.tex
\section{Introduction}

The rapid proliferation of large language models (LLM) has intensified the demand for scalable and performant inference infrastructures. 
However, the increasingly high computational requirements and large model sizes make it challenging to serve them in a cheap and efficient manner. 
Centralized AI services — dominated by hyperscale providers—create structural bottlenecks~\cite{bommasani2021opportunities,strubell2019energy}: they impose high operational cost~\cite{hoffmann2022training}, introduce single points of failure~\cite{bai2024beyond,chen2025electricity}, and fragmented GPU resources across data centers~\cite{gao2024lowgpu,weng2023fgd,10.1145/3488423.3519336,10.5555/3691938.3691968}. 
As model sizes grow and applications diversify, these limitations become 
untenable. 

A decentralized approach to LLM inference~\cite{petals,10.5555/3692070.3692951,helix,tong2025parallax} offers a compelling alternative by distributing computation across heterogeneous, independently operated nodes, enabling elastic capacity 
and improving robustness against single-point power outages. 
Together, these factors make decentralized AI not merely an architectural choice but a necessary evolution for building scalable, democratic, and highly-accessible AI systems~\cite{hui2025decentralization,khan2023decentralized}.

At the same time, the growing heterogeneity and fragmentation of compute resources across modern data centers and personal devices create an opportunity to enable the decentralized approach. 
Contemporary cloud environments routinely exhibit substantial pockets of underutilized GPUs and stranding effects caused by placement constraints~\cite{10.5555/3691938.3691968,10.1145/3488423.3519336}. 
In parallel, consumer hardware—ranging from high‑end laptops to gaming GPUs (e.g., NVIDIA RTX 5090) 
—has become powerful enough to execute meaningful slices of LLM inference. 
By aggregating idle capacity, exploiting locality, and dynamically routing computation across a diverse pool of contributors, decentralized LLM inference transforms what is currently wasted or siloed compute into a scalable, elastic, and cost‑effective inference fabric.


Decentralized LLM inference distributes transformer workloads across the public internet rather than within a data center’s high‑bandwidth fabric. Residential contributors in decentralized settings often operate at only \textcolor{checked}{20–500 Mbps}~\cite{fcc2024section706}, making cross‑node communication 500$\times$–2000$\times$ slower than typical intra-cluster links. As a result, network transfer—not heterogeneous computation—becomes the primary bottleneck: for example, sending a 5 MB activation tensor takes 25–200 ms on typical home uplinks, far exceeding the 1–3 ms compute time of a transformer block on a desktop-level GPU. This stark bandwidth gap underscores the importance of shifting the performance optimization target to be aggressively \textit{communication centric}. 

In this paper, we introduce a framework (named \name) and study how to optimize throughput of distributed inference of LLM at internet scales (characterized with low network bandwidth and heterogeneous compute), with a focus on communication optimizations.

The communication overhead has two dimensions: (1) the number of inter-node hops in the inference pipeline and (2) the volume of data each hop carries. The two dimensions are entangled, creating challenges to reduce or hide the communication overhead. In particular, in \name, we explore three optimization techniques: layer assignment, micro-batching, and tensor offloading. The layer assignment decides the placement of transformer layers across geo-distributed GPUs, which impacts the overhead (1); the micro-batch decomposes the input batch to smaller batches for scheduling, such that we can overlap communication and computation, which impacts the overhead (2); the tensor offloading uses CPU memory to hold GPU memory spill such that we can use less GPU, hence impacting the overhead (1). The effectiveness of the three techniques interacts with each other because of GPU memory capacity constraint and I/O overhead for tensor offloading (\S\ref{sec:design}). As a result, coordinating the three techniques to maximize inference throughput is challenging. 


\name addresses this problem by formulating the coordination of the three techniques as an optimization problem, and solves it using dynamic programming. Furthermore, the formulation considers the heterogeneity of GPU and network, and the solving process is lightweight because of the constrained formulation.

On top of the synergies of the three techniques, we further push up inference throughput by using lossless compression and speculative decoding. The lossless compression reduces the volume of activations communicated across the internet without impacting the inference accuracy. The existing lossless compression methods either lose floating point structures (e.g., exponent and mantissa) ~\cite{zstd,zlib} or focus on LLM weights~\cite{zipnn} (not activations). \textcolor{checked}{We characterize the values of activations,  and reveal that by re-organizing data layout of activations, we can expose more repeated patterns for the compression algorithm to tap and hence improve efficiency.} 

Speculative decoding~\cite{chen2023accelerating, liu2023online,specinfer} (SD) has been employed in centralized environments to improve the throughput of LLM inference. SD uses a lightweight draft-model to propose candidate tokens and a heavyweight target-model to parallelize the verification of candidate tokens. However, when using SD in a decentralized environment, we face a challenge from slow interconnect: the candidate tokens, when transferring from the draft model to the target model across the internet, substantially amplifies inter-node communication volume, causing SD to yield \textit{negative} throughput gains compared to standard autoregressive decoding. To make SD useful in the decentralized environment, we introduce a learned classifier placed right before the data communication. The classifier efficiently prunes the candidate tokens to reduce the communication volume while effectively maintaining the token acceptance rate at the target model. 

We evaluate \name with multiple network environments with diverse bandwidth using a series of LLM models. Compared to the baselines of state-of-the-art decentralized LLM inference systems (i.e., Helix~\cite{helix} and Petals~\cite{petals}), \name improves service throughput by \textcolor{checked}{up to 1.76$\times$} while reducing average latency by \textcolor{checked}{up to 43.20\%}. 

In summary, we make the following contributions.

\begin{itemize}[noitemsep,topsep=2pt,parsep=0pt,partopsep=0pt,leftmargin=1.5em]
    \item We recognize the communication as the major bottleneck for building performant LLM inference at internet scales, and introduce a framework with multi-dimensional communication optimization. 
    
    \item We use a spectrum of techniques to reduce or hide communication overhead in decentralized environment; we identify their interactions and formulate them into an optimization problem to maximize inference throughput. 
    
    \item We customize lossless compression and SD to accommodate unique features of communicating tensors and decentralized environment. 
     
\end{itemize}

%% file: text/background.tex
\section{Background}



\textbf{Decentralized AI.} Across geo-distributed GPUs, the pipeline parallelism is commonly employed in decentralized AI~\cite{petals,10.5555/3692070.3692951,helix,tong2025parallax,sailor,douillard2023diloco,ryabinin2023swarm,senghaasdiloco,NEURIPS2021_41a60377,jaghouar2024intellect1technicalreport}. This means that the LLM is partitioned across GPUs, each of which contains a group of consecutive layers of the LLM. During inference, the inference request proceeds stage by stage, and only the activations at the stage boundaries are transferred between consecutive stages. The pipeline parallelism is a practical approach to enable decentralized AI, because the communication to transfer activation tensors across two pipeline stages is point-to-point, and the communication volume is much smaller (compared with tensor model parallelism~\cite{narayanan2021efficient}, another common parallelism scheme for distributed inference), and can be managed to be $O(100KB)$ across two stages.  Tensor model parallelism is not suited for the internet-scale LLM inference, because it requires collective communications, such as \texttt{AllReduce}, within transformer layers~\cite{shoeybi2019megatron}. Such communication patterns are sensitive to latency and bandwidth, which in practice confines tensor model parallelism to single-node or other tightly coupled environments.


In the setting of decentralized AI, given the long latency at the internet scale, we focus on token throughput (not latency) and aim to serve offline LLM inference. There are many use cases of offline LLM inference, such as batch processing large document sets~\cite{zhang2025scaledoc},  automated report generation~\cite{walden2025auto}, local photo captioning or organization~\cite{patel2025alt}, and offline creative writing assistant~\cite{wang2024weaver}. For those use cases, the decentralized AI provides a cost-effective solution with maximized throughput and reasonable latency~\cite{wu2025deserve}.

\textbf{Tree-based speculative decoding (SD).} SD~\cite{chen2023accelerating} is an acceleration technique for LLM in which a lightweight \textit{draft model} rapidly proposes a short sequence of candidate next tokens—often accompanied by their probability distributions—and passes these proposed continuations to a larger, more accurate \textit{target model} for verification. The target model evaluates the draft tokens in parallel, accepting those that fall within its own high‑probability predictions and rejecting or correcting any that diverge. By allowing the target model to skip many decoding steps while still ensuring that the final output adheres to its distribution, SD substantially improves throughput without compromising output quality.

Tree-based SD~\cite{specinfer} extends SD by organizing draft tokens into a tree structure, where each path represents a candidate continuation. The target model verifies all paths in a single forward pass via tree attention masking, substantially increasing the hypothesis space evaluated per step. 


%% file: text/motivation.tex
\section{Motivation}
\textbf{Analysis on communication data path.} The communication data path at the internet scale across GPUs is fundamentally different from that at the data center scale. 

At the data center scale, inter-GPU communication across nodes can rely on GPUDirect RDMA. In particular, the data from a GPU memory is first moved through the local GPU interconnect (e.g., NVLink and PCIe) to the network interface. From there, it is handed off to a network adapter-using RDMA-capable technologies (e.g., Infiniband or RoCE) which transports data across data-center fabric with minimum CPU involvement. On the receiving node, the network interface card (NIC) delivers payloads directly into target GPU's memory space, enabling fast, low-latency communication. 

In the decentralized environment, the end‑to‑end GPUDirect RDMA is rarely possible across the open internet because (1) RDMA requires a lossless fabric, but internet routing introduces congestion, packet loss, and variable Maximum Transmission Unit (MTU), which conflict with RDMA; and (2) the security layers (TLS, IPsec, and QUIC) needed by internet are on the CPU (not on NIC or GPU). So in real‑world geo‑distributed GPU communication, CPU memory must be involved, at least for packetization, encryption, and transport handling, even if the GPU‑side data movement is optimized. 


\textbf{Communication performance.} We examine the role of communication in internet-scale LLM inference. We use LLaMA-30B with batch size 32 and sequence length 128. 
The model is evenly partitioned across three nodes, each with 20 layers on an NVIDIA RTX 5090 GPU, using pipeline parallelism. We use two environments: (1) W1: a data center connected by 45 Gbps Ethernet; 
and (2) W2: three nodes distributed in Maryland (MD), North Carolina (NC), and Pennsylvania (PA). We use Petals~\cite{petals} (a framework for internet-scale AI) for W2. \textcolor{checked}{Along the forward pipeline path, the bandwidth of inter-stage links is 331.0~Mbps from MD to NC and 76.5~Mbps from NC to PA}.  
To make the comparison fair, we use Petals for W1 too.

\begin{figure}[t]
    \centering
    \begin{subfigure}[t]{0.49\columnwidth}
        \centering
        \includegraphics[width=\linewidth]{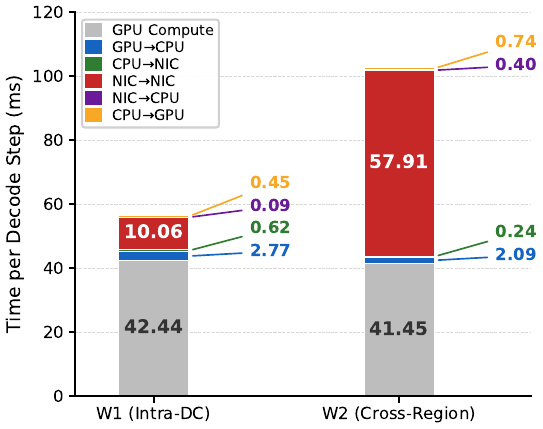}
        \caption{Per-step decoding-time breakdown.}
        \label{fig:motivation_comm_decoding}
    \end{subfigure}
    \hfill
    \begin{subfigure}[t]{0.49\columnwidth}
        \centering
        \includegraphics[width=\linewidth]{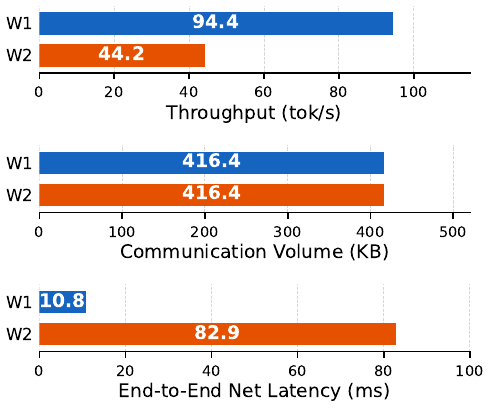}
        \caption{Throughput, communication volume, and end-to-end latency.}
        \label{fig:motivation_comm_thr}
    \end{subfigure}
    \caption{Performance comparison between W1 and W2.}
    \label{fig:motivation_comm}
\end{figure}

Figure~\ref{fig:motivation_comm} summarizes the communication behavior under W1 and W2. We break it down to GPU--CPU communication time ($T_{GPU \leftrightarrow CPU}$), CPU--NIC staging time ($T_{CPU \leftrightarrow NIC}$), NIC--NIC communication time ($T_{NIC \leftrightarrow NIC}$) exposed to the critical path, and per-stage GPU compute time. 

We have the following observations.

\textbf{Observation 1.}  Inter-site communication becomes the dominant component on the critical path once execution moves to the internet scale. 

With the same communication volume (416 KB per pipeline stage), $T_{NIC \to NIC}$ rises from 10.06~ms in W1 to 57.91~ms in W2. In W1, The communication is a minor cost (17.8\% of inference time); in W2, it reaches 1.4$\times$ GPU compute time (56.6\% of inference time).

\textbf{Observation 2.} The local staging overhead is small, compared to the communication time over the internet.

In W2, the host-mediated components are limited: $T_{GPU \to CPU}$, $T_{CPU \to NIC}$, $T_{NIC \to CPU}$, and $T_{CPU \to GPU}$ are 2.09~ms, 0.24~ms, 0.40~ms, and 0.24~ms, respectively. Their sum is only 2.97~ms, far below the 57.91~ms spent on $T_{NIC \to NIC}$. 
The main bottleneck is not local movement among GPU, CPU, and NIC. It is the transfer of activations across internet sites.

The above observations point to communication optimization rather than tuning of local computation as the main goal for improving internet-scale LLM inference. Since the communication dominates inference time, any computation (e.g., compression) that can save communication volume would be valuable. Observation 2 rules out local GPU--CPU and CPU--NIC staging as performance optimization targets.

%% file: text/overview.tex
\begin{figure*}[!t]   
    \centering
     \includegraphics[width=0.95\linewidth]{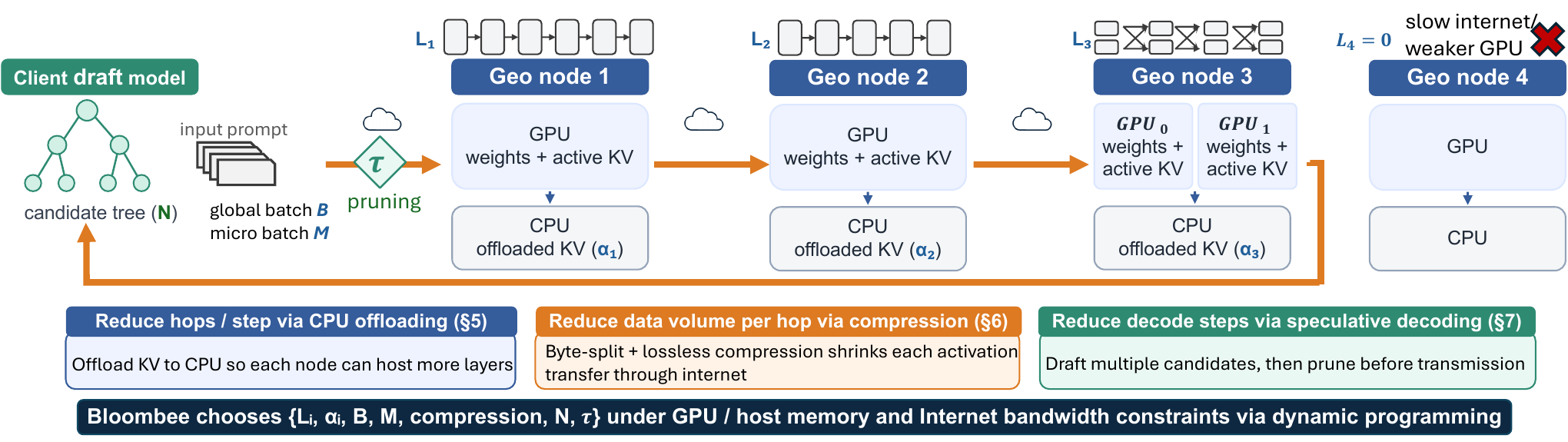}
     \vspace{-5pt}
    \caption{\name with heterogeneous GPU nodes connected over the internet form a pipeline-parallel inference system. \name jointly optimizes layer assignment ($L_i$), offloading($\alpha_i$), micro-batching($M$), compression, and speculative decoding ($\tau$ and $N$) to reduce communication overhead. Nodes with slow internet or weak GPUs that cannot contribute to system throughput are excluded from the pipeline ($L_i = 0$). }
    \vspace{-5pt}
    \label{fig:overview}
\end{figure*}

\section{\name Design Overview}
\label{sec:overview}

Driven by the observations, \name adopts a communication-centric design that jointly optimizes three factors: the number of inter-node hops per decoding step, the data volume per hop, and the total number of decoding steps, under GPU memory constraints, as shown in Figure~\ref{fig:overview}.

\textbf{Reducing hops.} \name consolidates more transformer blocks onto fewer nodes by adaptively offloading KV cache, which grows linearly with both batch size and sequence length, from GPU to host CPU memory when GPU capacity is insufficient. This frees GPU capacity for additional blocks and reduces pipeline hops. \name further decomposes each batch into micro-batches to overlap computation and communication across stages. These decisions are coupled through the GPU memory budget, and \name solves their coordination via dynamic programming  (\S\ref{sec:design}).

\textbf{Reducing volume.}  \name applies lossless compression to shrink the activation payload at each hop. The key insight is that byte lanes of serialized floating-point activations have markedly different entropy. A byte-split transform separates them before compression, yielding higher ratios than compressing the raw stream (\S\ref{sec:compression}).

\textbf{Reducing steps.} \name adopts speculative decoding to reduce the number of decoding steps per request, where a lightweight draft model on the client proposes candidate tokens and the target model deployed across \name's pipeline verifies them. Unlike single-machine settings, transmitting draft tokens for verification across geo-distributed nodes inflates communication overhead. \name addresses this through draft token pruning, communication-efficient batched transmission, and asynchronous KV cache management (\S\ref{sec:sd}).

%% file: text/design.tex
\section{Geo-Distributed High-Throughput Scheduling}
\label{sec:design}

\subsection{Throughput-Optimized Pipeline Planning}
\label{sec:stagetime}

\textbf{System setup.}
\name partitions an LLM of $L$ transformer blocks across $N$ heterogeneous nodes in a linear pipeline (Figure~\ref{fig:overview}). Between nodes, pipeline parallelism forwards activations over the internet; within each node, tensor parallelism splits layers across local GPUs. The scheduling granularity is the transformer block: modern LLMs stack identical blocks (self-attention followed by feed-forward), so each block contributes the same compute cost and the same KV cache footprint. 
The planner assigns each node a contiguous range of $L_i$ blocks, with $\sum_i L_i = L$, and $L_i \geq 0$. 

\textbf{Throughput objective.}
The pipeline processes a batch of $B$ requests, each generating $s$ tokens.
\begin{equation}
\max \; \frac{B \cdot s}{\sum_{i=1}^{N} \bigl( T_{\text{comp}}(i) + T_{\text{comm}}(i) \bigr)}
\end{equation}
where $T_{\text{comp}}(i)$ is the compute time at node $i$ and
$T_{\text{comm}}(i)$ is the communication time to the next node.

\textbf{Per-node compute time.}
As described in \S\ref{sec:overview}, \name offloads a fraction
$\alpha_i \in [0, 1]$ of each node's KV cache to host CPU memory,
splitting attention computation between CPU and GPU while MLP
remains on GPU. The per-stage compute time is:
\begin{equation}
\begin{split}
T_{\text{comp}}(i) &= L_i \cdot t_{\text{block}}(\alpha_i) \\
t_{\text{block}}(\alpha_i) &= t_{\text{mlp}} + (1 - \alpha_i) \cdot t_{\text{attn}}^{\text{gpu}} + \alpha_i \cdot t_{\text{attn}}^{\text{cpu}}
\end{split}
\end{equation}
All latencies are profiled per node. When a node contains multiple
GPUs, \name applies intra-node tensor parallelism; the profiled
latencies reflect this.

\textbf{Inter-node communication time.} Each stage boundary incurs a network hop over the public internet. The per-hop communication time is:
\begin{equation}
\small
T_{\text{comm}}(i) = \lambda_i + \frac{B \cdot d}{B_{\text{net}}(i)}
\end{equation} 
where $\lambda_i$ is the network propagation latency between node $i$ and node $i{+}1$, $d$ is the per-request activation size determined by the model's hidden dimension, and $B_{\text{net}}(i)$ is the effective link bandwidth.

\textbf{Constraints.} The layer assignment, offload ratio, and batch size are coupled through memory. Assigning more layers to a node increases weight memory, which forces a higher offload ratio $\alpha_i$ or a smaller batch size $B$. Formally, all blocks must be assigned ($\sum_i L_i = L$), each node's GPU must accommodate model weights, activation workspace, and the GPU-resident portion of KV cache, and each node's host memory must accommodate the offloaded KV cache.

\textbf{Decision variables.} For a given batch size $B$, the planner determines the layer assignment $\{L_1, \ldots, L_N\}$; the offload ratios $\{\alpha_i\}$ are then derived from the GPU memory constraint. Setting $L_i = 0$ excludes a node and its adjacent internet hops from the pipeline.

\textbf{Layer placement via dynamic programming.} The placement problem assigns $L$ blocks to $N$ nodes in pipeline order. Because each node holds a contiguous range of blocks, the problem reduces to an interval partition, which \name solves exactly via dynamic programming (DP) in $O(N \cdot L^2)$ time. For a pipeline of $N=8$ nodes and a model with $L=80$ blocks, the solver completes in under one millisecond. In contrast, Helix~\cite{helix} formulates placement over arbitrary routing graphs using mixed-integer linear programming, which requires up to hours of solving time with large number of nodes $N$.

For each candidate batch size $B$, the solver iterates over nodes in pipeline order, evaluating all feasible layer counts per node. For each candidate assignment, it computes $T_{\text{comp}}(i)$ and $T_{\text{comm}}(i)$ from a profiled cost table, derives $\alpha_i$ from the GPU memory constraint, and selects the partition that minimizes total pipeline time. The planner runs the solver for each candidate $B$ and selects the configuration with the highest throughput.

The DP captures heterogeneity without manually tuned placement rules: nodes with fast PCIe and large CPU memory host more layers because offloading overhead is low, nodes behind slow internet links are skipped to remove costly hops, and nodes with weaker GPUs
receive fewer layers to avoid becoming the bottleneck.

\subsection{Micro-batch Pipelining}
\label{sec:batch}

\begin{figure}[t]
    \centering
    \includegraphics[width=\linewidth]{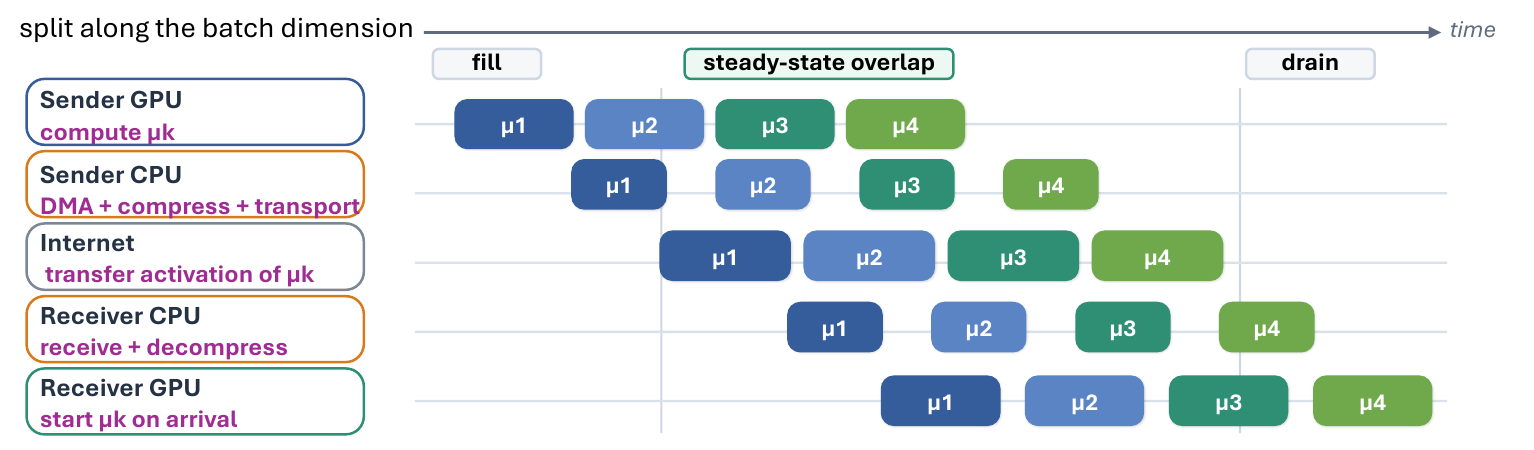}
    \vspace{-5pt}
\caption{Micro-batch pipelining in \name. 
After a node finishes computing $\mu_k$, the activation moves to CPU memory for compression and transport, while the sender GPU immediately starts $\mu_{k+1}$. In steady state, computation and communication proceed in parallel (Equation~\ref{eq:pipeline}). }
     \vspace{-5pt}
    \label{fig:pipeline}
\end{figure}

In a geo-distributed pipeline, each node's GPU is idle while its activation transfers over the internet. Since network transfer dominates stage time, GPU utilization is low. \name splits the batch of $B$ requests into $M$ micro-batches of size $b = B/M$, where each micro-batch contains a subset of requests (i.e., partitioned along the batch dimension) to overlap communication with computation, as shown in Figure~\ref{fig:pipeline}. When a node finishes computing a micro-batch, it copies the output activation to CPU memory via DMA, freeing the GPU to immediately begin the next micro-batch. 
The CPU then handles compression and network transmission in parallel. This overlap exploits the internet-scale data path: since GPU Direct RDMA is unavailable, activations must transit through CPU memory, which naturally decouples GPU computation from network transfer. As the pipeline fills, all stages operate concurrently on different micro-batches.

In steady state, the throughput objective with micro-batching becomes:
\begin{equation}
\begin{split}
\max\; &\frac{B \cdot s}
  {(M + N - 1) \cdot \max_{i}\, T_{\text{cycle}}(i)} ,\\
where\;\;\; T_{\text{cycle}}(i) &= \max\bigl(
  T_{\text{comp}}(i),\; T_{\text{comm}}(i) \bigr)
\end{split}
\label{eq:pipeline}
\end{equation}
The per-stage cost becomes the maximum of computation and communication, rather than their sum. The factor $(M + N - 1)$ accounts for pipeline fill and drain. As $M$ grows, the fill and drain overhead amortizes and throughput is dominated by the slowest stage. The planner selects $M$ jointly with $B$ by enumerating candidate values during the dynamic programming solver (\S\ref{sec:stagetime}). 

\section{Communication Compression}
\label{sec:compression}
We adopt lossless compression, instead of lossy compression (e.g., quantization). The tensors transferred are intermediate activations, rather than static model weights stored offline. This distinction is important: activations are generally more difficult to quantize than weights, in part due to outliers~\cite{smoothquant}. Prior work also shows that reducing activation precision can introduce noticeable accuracy degradation in the transient states used during inference~\cite{wkvquant}. Although some systems quantize transmitted activations to reduce latency~\cite{commcomp}, they alter the hidden states consumed by downstream pipeline stages. Lossless compression preserves tensor values bit-for-bit without introducing an accuracy–bandwidth tradeoff.


\textbf{Motivation of our design.}
Lossless compression employs two main techniques, repetition removal (created in LZ compression~\cite{10.1109/TIT.1977.1055714,10.1109/TIT.1978.1055934}) and entropy encoding (e.g., \cite{4051119,5391119}). LZ compressors detect repeated sequences of multiple bytes—usually four or more—and replace them with compact back-references to earlier occurrences, reducing storage requirements. Entropy coders, in contrast, analyze the probability distribution of individual bytes and shrink the data by encoding frequent symbols with shorter bit-level representations. Many widely used compressors, such as ZSTD~\cite{zstd} and zlib~\cite{zlib}, combine these two approaches by removing repeated sequences and then applying entropy coding. 

\begin{figure}[!t]
    \centering
    \includegraphics[width=\linewidth]{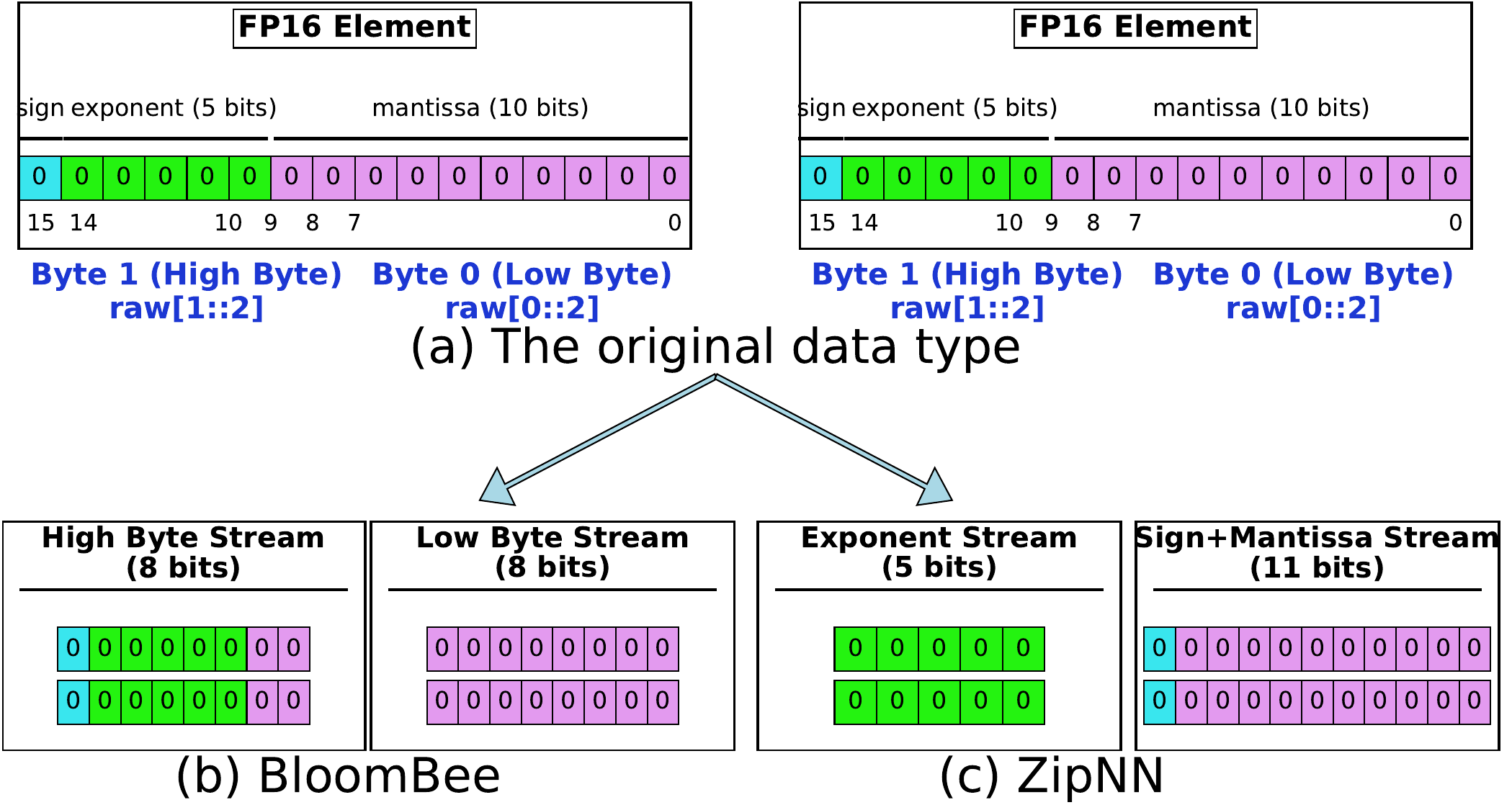}
    \caption{Comparison of data layouts.}
    \label{fig:fp16_byte_split_layout}
\end{figure}

We study two most popular lossless compressors, ZSTD and zlib and one state-of-the-art, ZipNN~\cite{zipnn} for \name. ZSTD and zlib serialize the input data into a byte stream and then detect repeated bytes within a sliding window over the stream. This method is general, but loses floating-point structures (e.g., mantissa and exponent) once the data values are serialized. ZipNN improves this method by separating exponent bits and mantissa bits (see Figure~\ref{fig:fp16_byte_split_layout}.c), which leads to more repeated patterns. ZipNN focuses on AI model weights.  

In the context of decentralized LLM inference, we need to compress \textit{activations} (not weights). Hence, we aim to answer the following two fundamental questions. 

\begin{itemize}[noitemsep,topsep=2pt,parsep=0pt,partopsep=0pt,leftmargin=1.5em]
    \item Does the separation of exponent bits and mantissa bits in activations reveal more compression opportunities?
    \item How should the separation happen to maximize compression effectiveness?
\end{itemize}

\begin{figure}[!t]
    \centering
    \includegraphics[width=\linewidth]{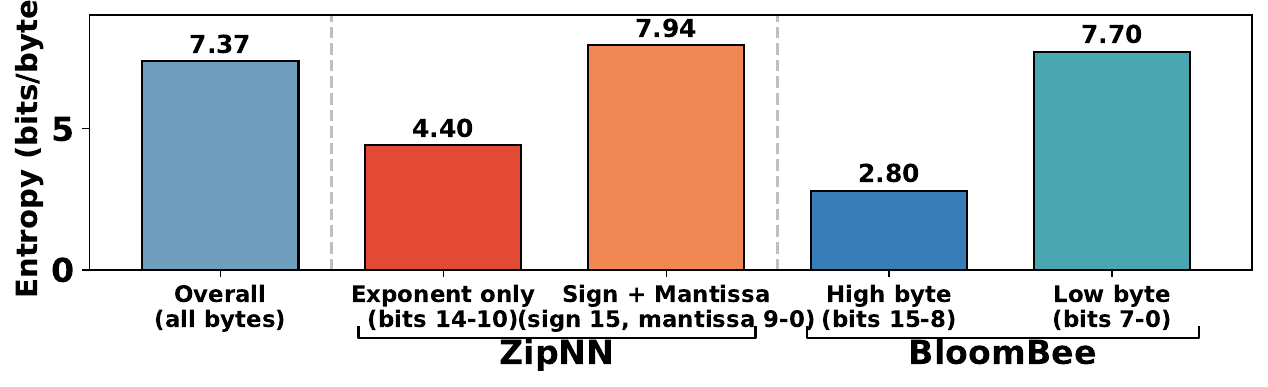}
    \vspace{-15pt}
    \caption{Entropy of the raw serialized FP16 byte stream and the component streams induced by ZipNN and \name.}
    \label{fig:fp16_hist_views}
\end{figure}

To answer these questions, we profile activations dumped from multiple models (LLaMA-13B/30B/65B, Mixtral-8$\times$7B, and Falcon-40B) during decoding. 
Since activations use FP16, we study FP16. 
Figure~\ref{fig:fp16_hist_views} shows the results for LLaMA-13B using the metric \textit{entropy}. Entropy in data compression~\cite{shannon1948} measures the average information content or unpredictability of data. 
Low-entropy data is highly redundant (e.g., repeated patterns) and compressible, while high-entropy (random) data is not. 

\textcolor{checked}{Figure~\ref{fig:fp16_hist_views} shows that the entropy of the whole 16 bits is 7.37 bits/byte. Using the method of ZipNN, the separated exponent bits have an entropy of 4.40 bits/byte, which is significantly (40.3\%) lower than the whole 16 bits, while the mantissa bits (plus one sign bit) have an entropy of 7.94 bits/byte, which is slightly higher (7.7\%).} Overall, the separation of exponent and mantissa can bring benefits, which answers the first question. 

Figure~\ref{fig:fp16_hist_views}.c shows the entropy of another bit-separation method. In this method, the separation happens at the byte binary: we report the entropy for the first byte and second byte of the 16 bits. The first byte (named \textit{high byte}) includes 1-bit sign, 5-bit exponent, and the most significant two bits of mantissa); the second byte (named low byte) includes the least significant eight bits of mantissa. Comparing   Figure~\ref{fig:fp16_hist_views}.b and Figure~\ref{fig:fp16_hist_views}.c, we see that the high and low bytes show lower entropy than exponent and mantissa (plus one sign), revealing more opportunities for compression.   

The rationale behind the improvement from the new separation method is that the most significant two bits of mantissa in activations show repetitiveness. Combining them with repetitive exponent bits, we create longer repetitive sequences, which leads to more effective lossless compression. 



\textcolor{checked}{We have the similar observations in LLaMA-30B, LLaMA-65B, Mixtra-8$\times$7B, and Falcon-40B}.




\textbf{Compression design in \name.} Based on the above discussions, \name introduces a lightweight compressor. Similar to ZSTD and zlib, \name serializes the data into a stream buffer, and applies entropy coding of ZSTD after compression. But different from ZSTD, zlib, ZipNN, \name considers the floating point structure and uses the new bit-separation method. 

\textbf{Effectiveness of compression in \name.} We compare \name with ZSTD and ZipNN in terms of compressed size in percentage (smaller is better) and compression time. We use activations collected from LLaMA-13B for compression. Table~\ref{tab:compression_direct_compare} reports the \textcolor{checked}{average} results per inference among \textcolor{checked}{16 inference requests from the dataset AlpacaEval \cite{alpaca_eval}}. In conclusion, \name leads to the smallest compressed size, 33\% and 35\% smaller than ZSTD and ZipNN respectively, while using similar compression time. \textcolor{checked}{We also note that ZSTD slightly outperforms ZipNN even though ZSTD does not consider the floating point structure. We attribute ZSTD's better performance to its effective entropy coding. \name uses ZSTD's entropy coding plus the new separation method, hence beating both ZSTD and ZipNN.}

\textcolor{checked}{At the internet scale, the compression time can be a small portion of communication time. For example, In Table~\ref{tab:compression_direct_compare}, given overall activation size of 76.5 MiB,  the communication time is 27,911ms with the bandwidth of 20 Mbps. For a network with higher bandwidth (e.g., 500 Mbps), the compression can be turned off when the compression cannot bring performance benefits.}

\begin{table}[!t]
\centering
\caption{Comparison of lossless compression methods}
\vspace{-5pt}
\label{tab:compression_direct_compare}
\resizebox{\columnwidth}{!}{
\begin{tabular}{lcc}
\toprule
\textbf{Method} & \textbf{Compressed Size (smaller is better)} & \textbf{Total Compression Time} \\
\midrule
ZSTD                & 69\% & 290.0 ms \\
ZipNN                 & 71\% & 330.2 ms \\
\name   & \textbf{46\%} & 300 ms \\
\bottomrule
\end{tabular}
}
\end{table}

\input text/design_sd

%% file: text/design_sd.tex
\section{Speculative Decoding over Internet}
\label{sec:sd}

In \name's pipeline, each decoding step requires a full round of communication across all geo distributed stages. Reducing the number of decoding steps therefore directly reduces internet communication. 
Speculative decoding offers this opportunity: a draft model on client side proposes multiple candidate tokens, and the target model deployed with \name geo-distributed verifies them in a single pass. In \name, however, the benefit of fewer target-model passes must be weighed against the cost of sending a larger candidate set over every internet hop.

\subsection{Speculative Decoding under Network Constraints}

\name's communication-centric design creates a tension with speculative decoding. As discussed in \S\ref{sec:batch}, \name batches $B$ requests together to amortize the cost of each internet hop. Under autoregressive decoding, each request contributes one token's hidden state per step, so a batched hop carries $B \cdot D$, where $D$ is the hidden-state size of a single token. Under speculative decoding, each request carries a draft tree with $N$ candidate token states, increasing the per-hop payload to $B \cdot N \cdot D$. The same batching that makes internet-scale inference practical therefore amplifies the communication overhead of speculation.

Over the internet, the larger payload traverses every distributed stage boundary, and the additional transfer time can exceed the latency saved by reducing the number of target-model passes. Whether speculation helps therefore depends on the candidate set size $N$, the average acceptance rate of tokens per speculative pass $a$, and the available bandwidth $S$. Here, $a$ denotes the average number of output tokens committed per speculative pass.

\begin{table}[!t]
\centering
\small
\caption{Notation used in the latency model.}
\vspace{-5pt}
\label{tab:sd-notation}
\begin{tabular}{cl}
\toprule
\textbf{Symbol} & \textbf{Definition} \\
\midrule
$L$               & Total tokens to generate \\
$D$               & Hidden-state payload per token (MB) \\
$S$               & Network bandwidth (MB/s) \\
$t_{\text{rtt}}$  & Fixed per-transfer latency (ms) \\
$t_{\text{comp}}$ & Per-worker-node compute time, autoregressive (ms) \\
$m$               & Compute time ratio, speculative vs.\ autoregressive \\
$c$               & Draft model compute time per step (ms) \\
$n$               & Number of worker nodes \\
$N$               & Draft tree size (nodes) \\
$a$               & Average accepted tokens per step \\
$S^*$             & Break-even bandwidth (MB/s) \\
\bottomrule
\end{tabular}
\vspace{-5pt}

\end{table}

Table~\ref{tab:sd-notation} summarizes the notation. We consider the end-to-end latency for generating $L$ output tokens across $n$ internet-distributed pipeline stages. For simplicity, the model considers one transformer block compute and communication per stage. Under autoregressive decoding, the total execution time is 
\vspace{-5pt}
\begin{equation}
\small
T_{\text{auto}} = L \cdot \left( n \cdot t_{\text{comp}}
  + \frac{n \cdot B \cdot D}{S} + n \cdot t_{\text{rtt}} \right)
\label{eq:auto}
\end{equation}
Each token in $L$ incurs compute, transfer of the batched hidden states, and round-trip latency at every stage. 

With speculative decoding, the target model performs only $L/a$ verification passes on average, but each pass carries $N$ candidate token states and requires $m \times$ the compute of
one autoregressive pass:
\begin{equation}
\small
T_{\text{spec}} = \frac{L}{a} \cdot \left( c + n \cdot m
  \cdot t_{\text{comp}} + \frac{n \cdot B \cdot N \cdot D}{S}
  + n \cdot t_{\text{rtt}} \right)
\label{eq:spec}
\end{equation}
where $c$ is the draft-model generation time per speculative
pass. Speculation helps only when $T_{\text{spec}} <
T_{\text{auto}}$, demonstrated as

\begin{equation}
    \small
    \underbrace{\frac{B \cdot (N - a) \cdot D}{S}}_{\text{data transfer penalty}} 
    + \underbrace{(m - a) \cdot t_{\text{comp}}}_{\text{compute penalty}}
    + \underbrace{(1 - a) \cdot t_{\text{rtt}}}_{\text{RTT saving}}
    + \underbrace{\frac{c}{n \cdot a}}_{\text{draft cost}} < 0
    \label{eq:condition}
\end{equation}

Rearranging yields the break-even bandwidth:
\begin{equation}
\small
S > S^* =
\frac{B \cdot (N - a) \cdot D}
{(a - 1) \cdot t_{\text{rtt}} + (a - m) \cdot t_{\text{comp}} - c/n}
\label{eq:sstar}
\end{equation}

Equation~\ref{eq:sstar} exposes why speculative decoding is challenging in the internet setting. The RTT term improves with $a$, but data transfer grows with both batch size and candidate set size. When the payload increase dominates the reduction in passes, speculation slows decoding.

\begin{figure}[t]
    \centering
    \includegraphics[width=1\linewidth]{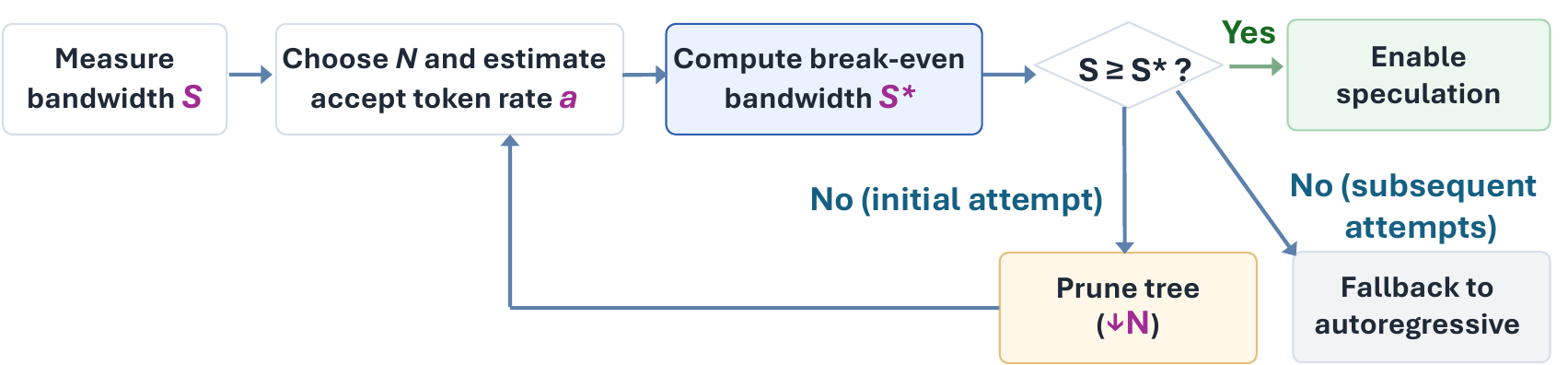}
    \vspace{-10pt}
    \caption{Bandwidth-aware SD configuration in \name.}
        \vspace{-10pt}
    \label{fig:sd_decision}
\end{figure}

To reduce the break-even bandwidth $S^*$, \name prunes low-probability candidates from the draft tree before data transfer, reducing $N$ while preserving most of the progress benefit (\S\ref{sec:pruning}). If the measured bandwidth remains below $S^*$ after pruning, \name disables peculation and falls back to autoregressive decoding, ensuring that speculation does not increase end-to-end latency. Figure~\ref{fig:sd_decision} demonstrates this decision process.

\subsection{Draft Tree Pruning and Efficient Verification}
\label{sec:pruning}

Equation~\ref{eq:spec} shows that the data transfer cost of speculation grows linearly with the number of candidate token states sent across each internet hop. Pruning therefore reduces communication directly. The challenge is that pruning can also remove candidates that would have survived verification, reducing the average acceptance ratio of $a$. \name addresses this tension with three mechanisms: early pruning at the first distributed stage, padding-free data transfer of the retained states, and asynchronous KV-cache compaction.

\textbf{Early pruning from first-stage hidden states.} A key observation is that the final-layer hidden states at the first worker node already encode enough context to predict whether a candidate is likely to survive full-model verification. \name therefore scores candidates before they cross the next internet hop.

Concretely, let $\mathbf{h}_i$ denote the hidden state of candidate token $v_i$ at the output of the first worker node. \name applies a lightweight proxy LM head $\hat{W}$ and obtains a local distribution estimate
\begin{equation}
\hat{p}_i = \mathrm{softmax}(\hat{W}\mathbf{h}_i).
\label{eq:proxy-head}
\end{equation}
The proxy head is trained offline to match the full-model distribution $p_i$ by minimizing the KL divergence $D_{\mathrm{KL}}(p_i \| \hat{p}_i)$. From $\hat{p}_i$, \name extracts three scalar features,
\begin{equation} 
\mathbf{f}_i = \left[\max(\hat{p}_i),\; \hat{p}_i[v_i],\;
\mathcal{H}(\hat{p}_i)\right],
\label{eq:features}
\end{equation}
which capture distribution peakedness, confidence in the candidate token, and overall uncertainty. A lightweight MLP maps $\mathbf{f}_i$ to a retention score $s_i \in [0,1]$, and \name drops candidate $v_i$ when $s_i < \tau$. The threshold $\tau$ controls the pruning ratio at inference time without retraining. The classifier is trained on the Alpaca dataset with binary labels indicating whether the candidate belongs to the full model's retained top-$K$ set, and adds only sub-millisecond overhead per decoding step.

\begin{figure}[!t]
    \centering
    \includegraphics[width=0.9\linewidth]{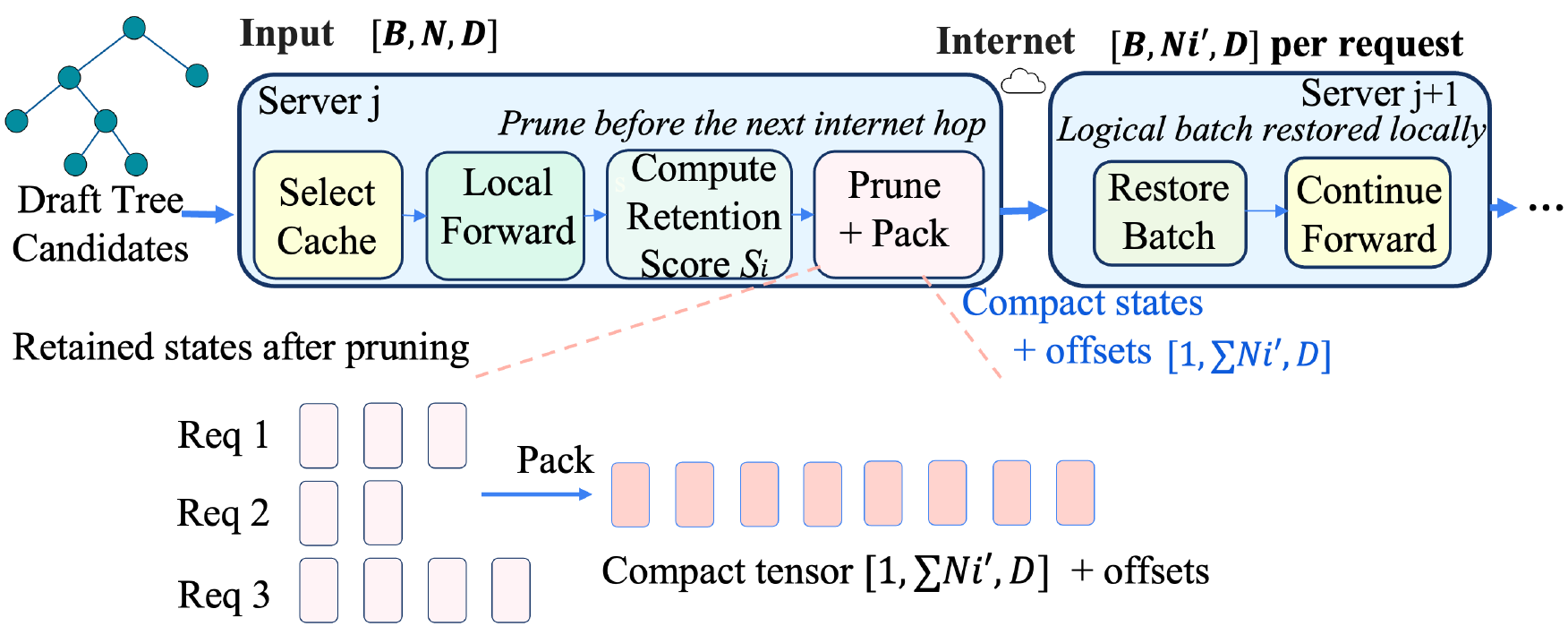}
    \vspace{-10pt}
    \caption{Data transfer for SD in \name.}
        \vspace{-5pt}
    \label{fig:transmission}
\end{figure}

\textbf{Padding-free data transfer.} After pruning, request $i$ retains $N'_i$ candidate token states, where $N'_i$ may differ across requests. The post-pruning batch is therefore ragged. Padding each request back to a common length would restore a regular layout, but it would also reintroduce the communication overhead that pruning eliminates. \name instead packs all retained states into a single contiguous tensor of shape $[1, \sum_i N'_i, D]$ and transmits it together with a small offset array that records per-request boundaries, as shown in Figure~\ref{fig:transmission}. The next worker node reconstructs the logical batch layout locally and continues the forward pass without extra network messages.

\textbf{Asynchronous KV-cache compaction.} Verification reveals which draft tokens were committed only after the client returns the accepted prefix at the beginning of the next decoding round. Eager compaction would therefore either block the current round or require an extra internet round-trip. \name instead piggybacks the acceptance metadata on the next inference request and defers compaction until that request arrives.

Each worker node keeps its KV cache in three regions: a compact prefix from earlier rounds, a hole region corresponding to previously rejected tokens, and a newly appended region for the current round. The hole region is masked during attention, so the cache remains immediately usable. Once the next request arrives, a background thread compacts the hole region in parallel with the current round's layer execution. 
In steady state, this overlap keeps compaction off the latency-critical path.

%% file: text/evaluation.tex
\section{Evaluation}
\label{sec:eva}

\textcolor{checked}{\name is an extension to Petals. Hence, in terms of computation (e.g., attention and MLP), \name and Petals have the same implementation. \name is open-sourced and has 31,216 lines of code in total.} \name has been deployed in multiple realistic use cases.

\subsection{Evaluation Setup}
\textbf{Baselines.} We compare \name against two baselines. Neither of them have any communication optimization techniques employed by \name.  (1) \textit{Petals}, an open-source framework for internet-scale LLM fine-tuning and inference. (2) \textit{Helix}, an LLM serving framework for heterogeneous GPU and network environments on the internet scale.

\textbf{Models.} We use LLaMA (13B, 30B, and 65B). 
Unless otherwise stated, the primary 
evaluation model is LLaMA-30B with batch size 32, as it is large enough 
to expose inter-stage communication overhead while still allowing
controlled evaluation across multiple network environments. We also evaluate Falcon-7B, Falcon-40B, and Mixtral-8$\times$7B; Their results are  reported in  Appendix~\ref{sec:appendix_transport}.

\textbf{Cluster setup.} We evaluate \name with three types of cluster setups:
(1) a single cluster (E1), (2) geo-distributed homogeneous clusters (E2-E5), and (3) geo-distributed heterogeneous clusters (E6).

E1 serves as a high-bandwidth (45 Gbps) environment without the internet. 
E2-E5 are four environments built upon E1 to emulate internet-scale inferences. We control interconnect bandwidth between nodes in E1, aligned with representative geo-distributed internet bandwidth~\cite{helix,petals,sailor}: the bandwidths for E2-E5 are 500 Mbps, 250 Mbps, 125 Mbps, and 20 Mbps respectively.

E6 is a realistic internet-scale environment with three GPUs: one NVIDIA A100 in California, one NVIDIA RTX~4090 in New Jersey, and one NVIDIA RTX~4090 in Canada. Table~\ref{tab:wan_bandwidth} summarizes the bandwidth between nodes in E6.

In our evaluation, we use single-GPU nodes in most evaluations, because multi-GPU nodes are harder to allocate consistently in the cloud~\cite{thorpe2023bamboo}. \name also supports tensor parallelism across GPUs within the same node when multi-GPU nodes are available.

\begin{table}[t]
\centering
\small
\caption{Bandwidths (Mbps) in E6.}
\label{tab:wan_bandwidth}
\setlength{\tabcolsep}{4pt}
\renewcommand{\arraystretch}{1.05}
\begin{tabular}{lccc}
\toprule
\textbf{Sender $\rightarrow$ Receiver}
& \textbf{California} & \textbf{New Jersey} & \textbf{Canada} \\
\midrule
California & --- & 312 & 280 \\
New Jersey & 347 & --- & 643 \\
Canada     & 305 & 577 & --- \\
\bottomrule
\end{tabular}
\end{table}

\subsection{Overall Performance}

\begin{figure*}[!t]
    \centering
    \includegraphics[width=\textwidth]{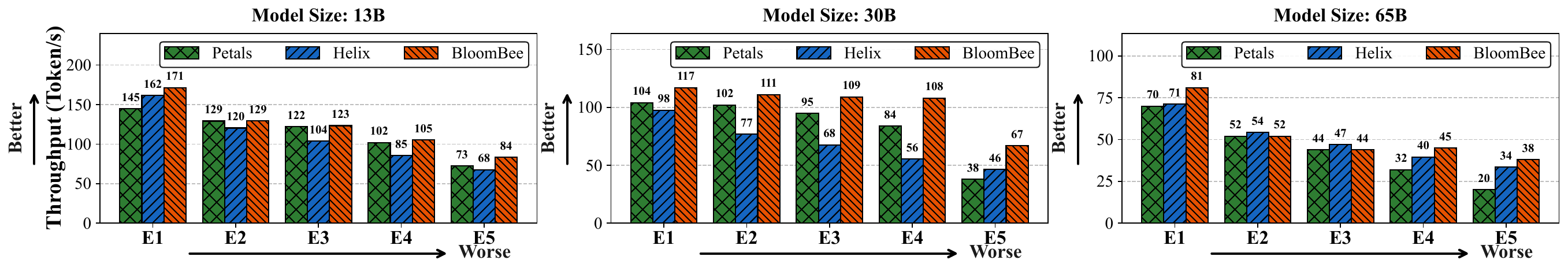}
    \vspace{-15pt}
    \caption{Overall performance across E1--E5 for various model sizes. }
    \vspace{-10pt}
    \label{fig:overall_performance}
\end{figure*}

\begin{table}[!t]
\centering
\caption{Best-performing communication optimization techniques in each environment.}
\vspace{-8pt}
\label{tab:env_best_perf_bs32}
\footnotesize
\setlength{\tabcolsep}{3pt}
\renewcommand{\arraystretch}{0.97}
\begin{tabular}{p{0.11\columnwidth} p{0.40\columnwidth} p{0.45\columnwidth} }
\toprule
\textbf{Model} & \textbf{Environments} & \textbf{Best techniques}  \\
\midrule
13B & E1 (2 nodes, 2 5090 GPUs)            & SD        \\
13B & E2,E3,E4 \tiny{(2 nodes, 2 5090 GPUs})            & Compression                   \\
13B & E5 (2 nodes, 2 5090 GPUs)            & Compression + micro-batching  \\
\midrule
30B & E1\tiny{(3 nodes, 3 5090 GPUs})             & SD          \\
30B &  E2 \tiny{(3 nodes, 3 5090 GPUs)}            & Autoregressive                \\
30B &  E3,E4,E5 \tiny{(3 nodes, 3 5090 GPUs)}            & Compression + micro-batching                \\
30B & E6 \tiny{(3 nodes, 2 5090 + 1 A100})            & Compression + micro-batching                \\
\midrule
65B & E1 (4 nodes, 8 5090 GPUs)            & SD               \\
65B & E2,E3 (4 nodes, 8 5090 GPUs)            & Autoregressive                \\
65B & E4,E5 (4 nodes, 8 5090 GPUs)            & Compression + micro-batching                \\
\bottomrule
\end{tabular}

\raggedright
\footnotesize
\textit{Autoregressive}: no compression, no micro-batching, no SD, and no offloading. 
\end{table}

We evaluate \name in E1-E5 against Petals and Helix. Figure~\ref{fig:overall_performance} shows the overall results.

\name outperforms both Petals and Helix across most environments. For example, in E5 with LLaMA-30B, \name achieves 67\,tok/s, demonstrating significant speedups over Petals ($1.76\times$) and Helix ($1.46\times$). In E3 and E4, \name matches or outperforms Petals while consistently outperforming Helix by notable margins.

The throughput gains are driven by two complementary effects: pruned SD reduces inter-stage communication volume by 60\% while preserving 96\% of the acceptance rate in high-bandwidth environments, and micro-batching with compression reduces communication overhead in bandwidth-limited environments.

 As observed in Figure~\ref{fig:overall_performance}, when we change the model size from 13B to 65B, the throughput gap between \name and Helix narrows. This is because of the change in number of layers per GPU. When the number of layers per GPU is larger, the computation per GPU becomes larger, which provides more opportunities to pipeline parallelism to hide communication overhead. Hence, there is less room for \name to optimize communication  performance.

Table~\ref{tab:env_best_perf_bs32} summarizes the communication optimization techniques selected by \name across various model sizes and evaluation environments. In the highest-bandwidth environment (E1), pruned SD is universally the primary driver: sufficient bandwidth allows SD to reduce the number of decoding steps while pruning makes transmission volume manageable.

As bandwidth drops (E2--E4), SD becomes less effective and is disabled. \name therefore disables SD. The model 13B solely relies on compression. For the model 30B, standard autoregressive decoding performs best in E2, while compression with micro-batching becomes the best choice in E3 and E4. For the model 65B in E2 and E3, standard autoregressive decoding performs best; the system overhead of compression and SD exceeds the communication savings at these environments.

Under severe bandwidth constraints (e.g., E5 for all model sizes and E4 for 65B), \name consistently applies compression and micro-batching. This configuration reduces per-step communication volume and hides transmission latency behind computation.





\subsection{Evaluation of Offloading}

We use LLaMA-30B with batch size 32 and sequence length 128, with compression, micro-batching, and speculative decoding disabled. With offloading, \textcolor{checked}{as selected automatically by \name,} all KV cache are placed in CPU, and \name can use just 2 GPUs, reducing the GPU count by 1. Both Helix and Petals have to use 3 GPUs. \textcolor{checked}{According to common AI-specific cloud vendors (Yotta~\cite{yottalabs} and VAST~\cite{vastai})}, we price each GPU at \$0.56 per hour, giving a 3-GPU baseline cost of \$1.68/h and a 2-GPU cost of \$1.12/h, and report throughput per GPU dollar, which considers the impact of offloading on both throughput and production cost.

\begin{figure}[!t]
    \centering
    \includegraphics[width=0.98\linewidth]{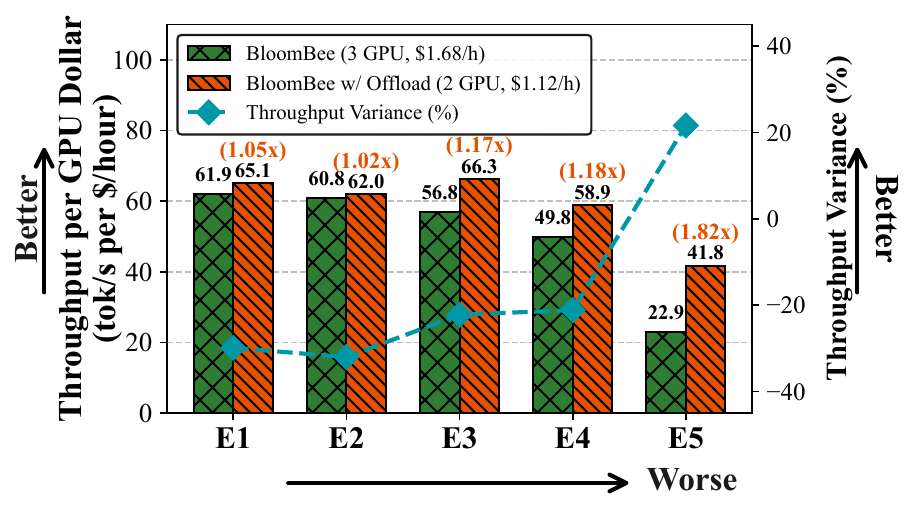}
    \vspace{-5pt}
    \caption{Cost efficiency (throughput per GPU dollar) with and without offloading.}
    \vspace{-5pt}
    \label{fig:offload_cost_efficiency}
\end{figure}

See Figure~\ref{fig:offload_cost_efficiency}. 
Offloading consistently improves cost efficiency across all 
environments, with the largest gains in bandwidth-limited settings. In E5, offloading achieves 1.82$\times$ higher throughput 
per GPU dollar (41.8 vs.\ 22.9\,tok/s/\$/h). In E3 and E4, the improvement is 1.17$\times$ and 1.18$\times$ respectively. Even in E1 and E2 with high-bandwidth, offloading still provides a minor gain (1.02$\times$ and 1.05$\times$). \textcolor{checked}{Note that with improvement of cost efficiency, no matter how small the improvement is, longer LLM service time will continuously bring benefits in larger saving of production cost.}

Figure~\ref{fig:offload_cost_efficiency} also shows the variance of throughput after using offloading (see the second y axis). We notice that with offloading, \name even \textcolor{checked}{brings 21.6\% improvement in throughput in E5}, because of saving in communication hops.

\textcolor{checked}{In another evaluation at E5, we use three GPUs and change the batch size. Using offloading, \name puts 30\% of KV cache to CPU, which enables larger batch sizes (up to 128), compared to no offloading with a smaller batch size 32. As a result, offloading brings 14.4\% improvement in throughput.}

\subsection{Evaluation of Micro-batching}

We use LLaMA-30B with batch size 32 and sequence length 128, with compression, offloading, and SD disabled. We use micro-batch size 16 unless otherwise stated.

\begin{figure}[!t]
        \centering
    \includegraphics[width=0.99\linewidth]{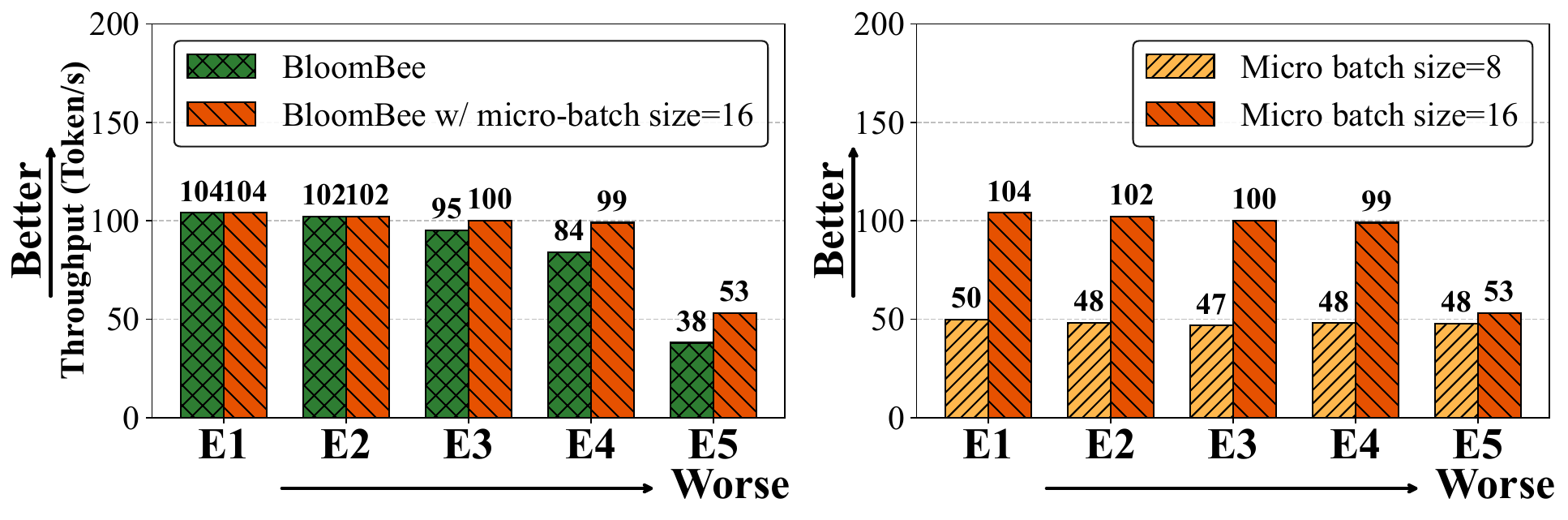}
    \vspace{-5pt}
    \caption{Evaluation of micro-batching.}
    \label{fig:microbatch_overview}
\end{figure}

See Figure~\ref{fig:microbatch_overview}. 
In E1 and E2 with relatively high bandwidth,  \name disables micro-batching automatically, and hence performs similarly to the case without micro-batching. From E3 onward, the gain of using micro-batching grows  monotonically:  +5.3\% in E3, +17.9\% in E4, 
and +39.5\% in E5.

We evaluate the impact of micro-batch size. The size 16 consistently outperforms the size 8 across all environments, with the size 8 plateauing at 47--50\,tok/s while size 16 matches or exceeds the no-micro-batching baseline in E1--E2 and substantially outperforms it in E3--E5. The larger micro-batches 
provide more opportunities for overlap, as the compute time per micro-batch grows

\subsection{Evaluation of Compression}

\begin{figure}[!t]
    \centering
    \includegraphics[width=\linewidth]{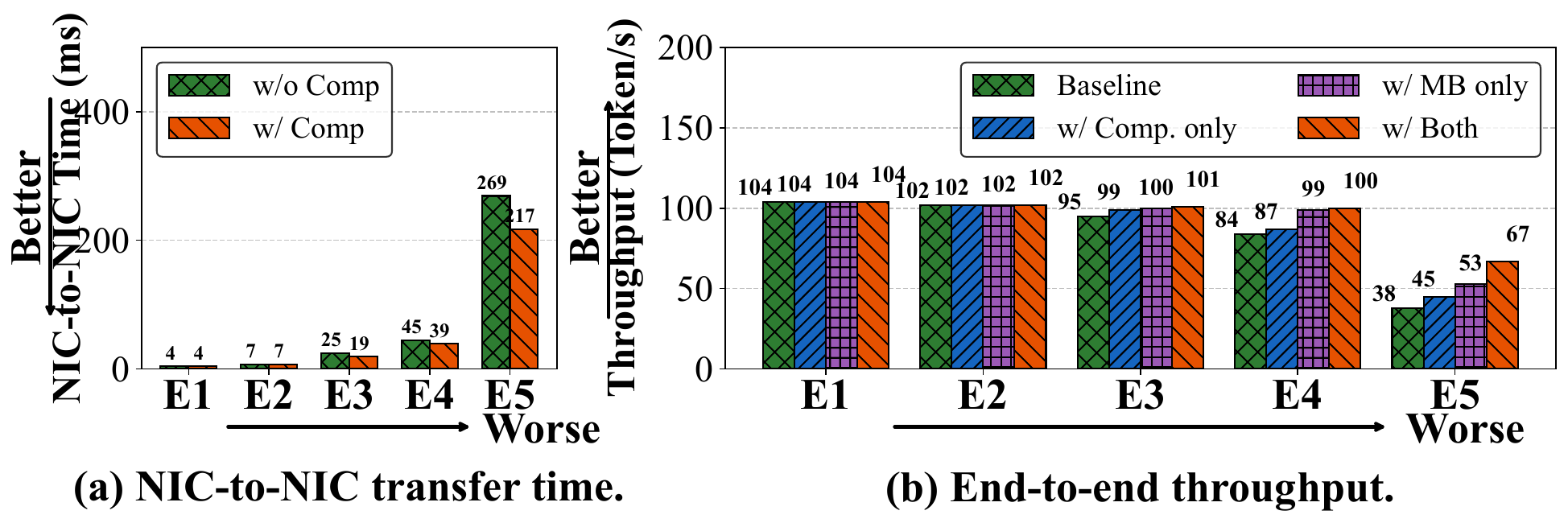}
    \caption{Compression evaluation with \name. ``MB'' = ``micro-batching''. ``Comp'' = ``compression''. } 
    \label{fig:compression_overview}
\end{figure}

We use LLaMA-30B with batch size 32 and sequence length 128. With this configuration, \name reduces the activation payload per hop from 416.4\,KB to 312.6\,KB. Since micro-batching can hide communication overhead and hence impacts the effectiveness of compression, we selectively add micro-batching for  evaluation. We compare four configurations: \name without compression and micro-batching (baseline), with compression only, with micro-batching only, and with compression plus micro-batching. See Figure~\ref{fig:compression_overview}.

The NIC-to-NIC transfer time reduction from compression grows with network constraint: negligible in E1 (4 to 4\,ms) and E2 (7 to 7\,ms),  but substantial in E3 (25 to 19\,ms), E4 (45 to 39\,ms), and E5  (269 to 217\,ms). This pattern drives the end-to-end throughput results.  In E1 and E2, all four configurations perform similarly, and \textcolor{checked}{\name disables compression since there is no benefit}.

In E5, where network transmission dominates, compression alone improves throughput by 18.4\%, micro-batching alone by 39.5\%, and the combination by 76.3\%. We note that the improvement of the combination is larger than the improvement summation of compression and micro-batching (i.e., 13.2\% + 39.5\%). This is because the micro-batching reshapes the data to be transferred across hops and accidentally brings more opportunities for effective compression.

\subsection{Evaluation of Speculative Decoding}
\suppressfloats[t]

We use LLaMA-30B with the batch size 32 and 
\textcolor{checked}{128 output tokens}. \textcolor{checked}{To reduce performance  variance because of prompt-dependent generation lengths, we randomly sample 
10 batches from the dataset Alpaca~\cite{alpaca}} and report mean throughput across runs. We compare three configurations: Helix, \name without any communication optimization (named \textit{Auto} and the baseline), unpruned SD (named \textit{SD}), 
and SD with MLP-based pruning (named \textit{SD+Prune}). 

Pruned SD yields consistent throughput improvements over the baseline 
in high-bandwidth environments: +12.5\% in E1 (117 vs.\ 104\,tok/s) 
and +1.0\% in E2 (103 vs.\ 102\,tok/s). In E3, pruned SD falls 
slightly below the baseline (75 vs.\ 95\,tok/s), as the transmission 
overhead of the draft tree begins to outweigh the reduction in decoding 
steps at this bandwidth. Unpruned SD performs worse than pruned SD in 
all three environments, confirming that pruning is essential to 
realizing the benefit of SD under network constraints: by reducing the 
transmitted tree size by 60\% while preserving 96\% of the acceptance 
rate, pruning shifts the operating point closer to the break-even 
bandwidth derived in \S\ref{sec:sd}. 

\begin{figure*}[!htbp]
    \centering
    \begin{subfigure}[b]{0.31\linewidth}
        \includegraphics[width=\linewidth]{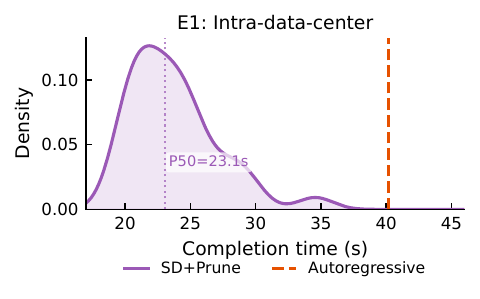}
        \caption{E1 (Intra-data-center) }
    \end{subfigure}
    \hfill
    \begin{subfigure}[b]{0.31\linewidth}
        \includegraphics[width=\linewidth]{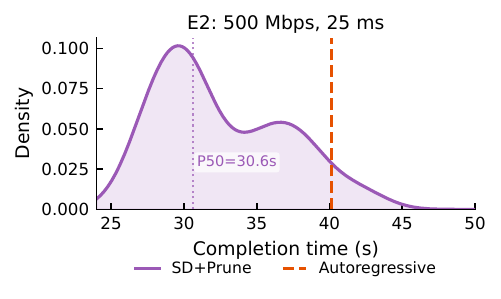}
        \caption{E2 (500\,Mbps, 25\,ms)}
    \end{subfigure}
    \hfill
    \begin{subfigure}[b]{0.31\linewidth}
        \includegraphics[width=\linewidth]{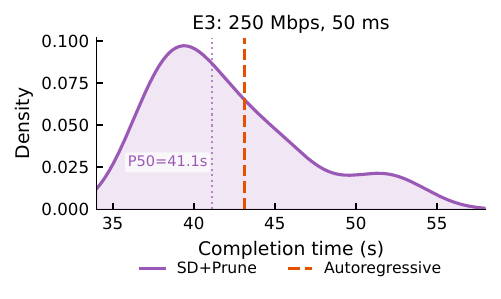}
        \caption{E3 (250\,Mbps, 50\,ms)}
    \end{subfigure}
    \caption{Distribution of per-sample completion latency. The dashed vertical line indicates the wall-clock time of 
    autoregressive (baseline) batch, at which all samples complete simultaneously. The dotted vertical line marks the 
    P50 of \textit{SD+Prune}. SD spreads 
    completions across a wide range, with the majority of samples finished well before the autoregressive baseline.} 
    \label{fig:latency_dist}
\end{figure*}

We compare the \textit{peak} throughput of \name using SD with the steady-state throughput \textcolor{checked}{(which is also the peak one)} of Helix. This comparison reflects the best case of \name against that of 
Helix. 
\textcolor{checked}{\name with pruned SD achieves 1.56$\times$, 1.46$\times$, and 
1.27$\times$ higher throughput than Helix in E1--E3 (152.73 
vs.\ 97.6, 112.17 vs.\ 76.9, and 85.82 vs.\ 67.6\,tok/s 
respectively) because of effectiveness of SD.}

Throughput alone, however, understates the benefit of SD in batched inference. Under autoregressive decoding, all samples in a batch complete simultaneously at the wall-clock time, since each step advances every sequence by exactly one token. SD breaks this uniformity: sequences that align  well with the draft model accept more tokens per step and finish 
earlier, while others are held until the batch tail.

We evaluate the impact of SD in \name on per-sample completion time. Figure~\ref{fig:latency_dist} shows the results. The per-sample completion times of 
\textit{SD+Prune} spread over a wide range, whereas \textit{Auto} 
appears as a single vertical bar. In E1, the median sample completes at 22.8\,s---43\% faster than the autoregressive baseline of 40.2\,s.  In E2, the median completion is 30.6\,s versus 40.3\,s ($-$24\%). In 
E3, the median is 41.1\,s versus 43.1\,s ($-$4\%). For LLM serving systems that dispatch completed requests as soon as they finish rather than waiting for the full batch to complete, this reduction in per-sample latency translates directly into lower user-perceived response time---a benefit 
orthogonal to batch throughput.

\subsection{Heterogeneous Clusters across Internet}
\suppressfloats[t]

\begin{figure}[t]
\centering
\includegraphics[width=\linewidth]{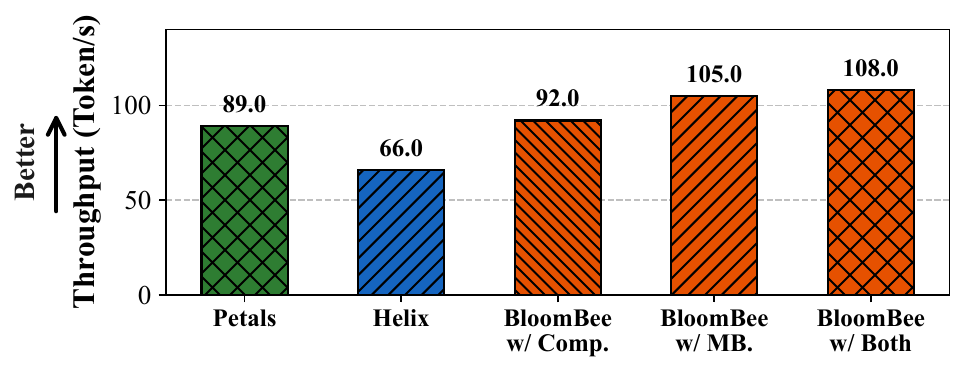}
\vspace{-3pt}
\caption{Throughput evaluation on realistic heterogeneous clusters across internet. ``MB'' = ``micro-batching''. ``Comp'' = ``compression''.}
\label{fig:e6_overview}
\end{figure}

We use LLaMA-30B with batch size 32 and sequence length
128. \name achieves the highest end-to-end throughput among all compared systems, shown in Figure~\ref{fig:e6_overview}. Petals reaches 89 tok/s, and Helix reaches 66 tok/s. BloomBee with compression alone further increases throughput to 92 tok/s. \name with micro-batching delivers a much larger gain, reaching 105 tok/s. When both optimizations are enabled, \name achieves the best result at 108 tok/s, outperforming Petals by 21.3\% and Helix by 63.6\%. The result shows that micro-batching is the main source of improvement in this setting, and compression brings a smaller but positive gain on top of it.

%% file: text/related_work.tex
\section{Related Work} 
\label{sec:related}


\textbf{LLM inference on geo-distributed GPUs.}
There are existing efforts using geo-distributed GPUs for LLM inferences with heterogeneous clusters and networks. Petals~\cite{petals} focuses on a pipeline parallel setup, and employs a greedy algorithm for model allocation and request scheduling in dynamic environments. Petals reduces communication overhead via quantization of weights and activations. HexGen~\cite{10.5555/3692070.3692951} is a concurrent work on LLM serving but works only for fixed pipelines. Helix~\cite{helix} studies model placement and scheduling based on Max Flow formulation and flexible per-request pipeline. Parallax~\cite{tong2025parallax} introduces dynamic pipeline construction with a two-phase scheduling strategy. This strategy creates replicas of pipeline stages and stitches layers from different replicas 
to balance load and improve utilization.

\textbf{Distributed LLM training on geo-distributed GPUs.} 
Sailor~\cite{sailor} co-optimizes resource allocation and 3D parallelization plans for distributed training over dynamic, heterogeneous, and geo-distributed clusters, emphasizing fast planning via accurate simulation and search-space pruning.  DiLoCo~\cite{douillard2023diloco} focuses on data parallelism for distributed training and proposes a new distributed optimizer to reduce frequency of collective communication. SWARM~\cite{ryabinin2023swarm} (and the extension of SWARM based on DiLoCo~\cite{senghaasdiloco})  constructs stochastic training-pipelines dynamically. Each node forwards activations or gradients to a randomly selected node in the next pipeline stage,  with probability proportional to the node’s throughput. This dynamically constructed pipeline enables dynamic load balance and fault tolerance. 
To reduce collective communication overhead, SPARTA~\cite{exo_sparta} only communicates a small random subset of gradients at each step. DeDLOC~\cite{NEURIPS2021_41a60377} (and DeDLOC extension, DiLoCo-FSDP2~\cite{jaghouar2024intellect1technicalreport} with \texttt{int8} all-reduce) use delayed parameter update during training to allow nodes to communicate less frequently and accumulate gradients at each node's own pace.  

%% file: text/conclusion.tex
\section{Conclusions}
In this work, we study LLM inference performance at internet scales, and customize the inference system design for low-bandwidth environments. Different from existing decentralized system designs, our study puts communication optimization as a first-class citizen. Our work sheds lights on how a decentralized LLM system can be designed to maximize throughput. 


%% file: text/appendix.tex
\section{Appendix}
\subsection{Specification-driven code generation.}

Integrating new model architectures (e.g., LLaMA) into \name requires writing a substantial amount of boilerplate code to match the runtime’s expected model interfaces and configuration formats. To reduce this effort, we adopt a specification-driven code generation approach based on structured model templates, depicted in Figure~\ref{fig:template-architecture}. 

The model developers provide a \textit{YAML specification} that describes model-specific parameters (e.g., layer structure and configuration fields). This specification is used to instantiate a set of Jinja2 \textit{templates} that generate boilerplate model scaffolding, including \texttt{block.py} (stage wrappers), \texttt{config.py} (model configuration), and \texttt{model.py} (model composition), with a consistent structure. The generated code covers repetitive scaffolding such as class definitions and configuration handling, while core computation logic (e.g., forward passes) remains manually implemented. The specification follows a constrained schema, allowing simple structural validation before code generation.

This approach simplifies model onboarding and reduces repetitive engineering effort, without changing the underlying execution semantics.

We evaluate spec-driven code generation against manual integration in deployment workflows. As shown in Table~\ref{tab:spec_codegen}, the approach reduces model integration time by 85\% and code review effort by 70\%, by eliminating boilerplate and enforcing interface consistency. Deployment error rates decrease by 95\%, indicating improved reliability from standardized generation. Automation further yields up to 3$\times$ faster end-to-end iteration, as components are regenerated directly from specifications. These results demonstrate improved efficiency and robustness in LLM deployment pipelines, with specification overhead amortized across repeated deployments.

\begin{figure}[!htbp]
\centering
\begin{tikzpicture}[
node distance=1.8cm,
box/.style={
    draw,
    rounded corners,
    align=center,
    minimum width=2.2cm,
    minimum height=0.9cm,
    line width=0.3pt
},
arrow/.style={->, thick}
]

\node[box, fill=blue!15] (spec) {Model\\Specification\\(YAML)};
\node[box, fill=yellow!20, right of=spec, xshift=1.1cm] (engine) {Template-based \\Generator};
\node[box, fill=orange!20, right of=engine, xshift=1.1cm] (modules) {Generated\\Infrastructure};

\node[box, fill=purple!15, below of=modules] (runtime) {Distributed\\Inference Runtime};

\draw[arrow] (spec) -- (engine);
\draw[arrow] (engine) -- (modules);
\draw[arrow] (modules) -- (runtime);

\node[below of=spec, yshift=0.7cm] {\footnotesize architecture + features};
\node[below of=engine, yshift=0.7cm] {\footnotesize deterministic synthesis};
\node[below of=modules, yshift=0.7cm] {\footnotesize block / config / model};

\end{tikzpicture}

\caption{Specification-driven code generation}
\label{fig:template-architecture}
\end{figure}
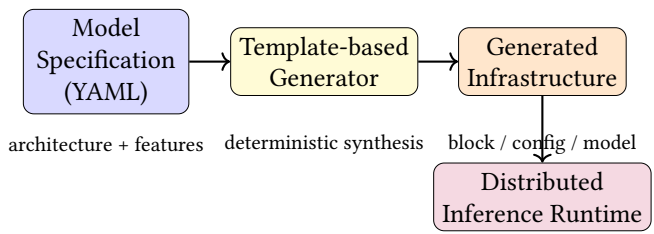
\begin{table}[!t]
\centering
\small
\caption{Impact of spec-driven code generation on development efficiency and reliability. Results compare template-driven integration with manual workflows.}
\label{tab:spec_codegen}
\begin{tabular}{|p{0.24\textwidth}|c|}
\hline
\textbf{Metric} & \textbf{Improvement} \\
\hline
Model integration time     & 85\% reduction \\ \hline
Code review effort         & 70\% reduction \\ \hline
Deployment error rate      & 95\% reduction \\ \hline
End-to-end iteration time  & 3$\times$ speedup \\ \hline
Manual integration effort  & Significantly reduced \\  \hline
\end{tabular}
\end{table}

\textbf{Example.} 
Listing~\ref{lst:model-spec} shows a high-level specification for a LLaMA-3-70B Transformer block, while Listing~\ref{lst:generated-block} presents the generated implementation. The generation process establishes a structured correspondence between specification fields and code components. Module-level entries such as \texttt{attention\_class}, \texttt{mlp\_class}, and \texttt{norm\_class} are mapped to concrete module instantiations in the constructor. Feature flags including \texttt{rotary\_embeddings}, \texttt{hf\_causal\_mask}, and \texttt{use\_cache} are compiled into conditional branches in the \texttt{forward} function, controlling execution behavior. Through this mapping, the generated code is directly driven by the specification rather than manually implemented. This ensures that architectural choices and feature configurations expressed in the specification are consistently and faithfully reflected in the resulting implementation, enabling interpretability and verifiability.

\lstdefinestyle{mystyle}{
  basicstyle=\ttfamily\footnotesize,
  frame=single,
  numbers=left,
  numberstyle=\tiny\color{gray},
  breaklines=true,
  tabsize=2,
  keywordstyle=\color{blue!60!black},
  commentstyle=\color{green!40!black},
  stringstyle=\color{orange!60!black},
  backgroundcolor=\color{gray!5}
}
\lstset{style=mystyle}
\begin{lstlisting}[
language=yaml,
caption={Model specification for LLaMA-3-70B},
label=lst:model-spec,
basicstyle=\ttfamily\footnotesize,
frame=single,
numbers=left,
numberstyle=\tiny\color{gray},
breaklines=true,
tabsize=2
]
model_name: "llama-3-70b"
attention_class: "FlexGenAttention"
mlp_class: "FlexGenMLP"
norm_class: "RMSNorm"

features:
rotary_embeddings: true
hf_causal_mask: true
kv_cache_reorder: false
use_cache: true

shared_impl_snippet: "path/to/custom_forward.py"

distributed:
dht_prefix: "llama3-70b"

\end{lstlisting}

\subsection{Transport Breakdown for Additional Model Architectures}
\label{sec:appendix_transport}

We report baseline transport breakdowns for Falcon-7B, Mixtral-8$\times$7B, and Falcon-40B across E1--E5 to assess whether \name's communication profile generalizes beyond LLaMA. Tables~\ref{tab:transport_breakdown_falcon7b_bs32}, \ref{tab:transport_breakdown_mixtral_bs32}, and \ref{tab:transport_breakdown_falcon40b_bs32} report the detailed transport breakdowns for Falcon-7B, Mixtral-8$\times$7B, and Falcon-40B, respectively. Across model families, scales, and GPU counts, the results confirm the same communication-dominated behavior observed in the main evaluation.
\begin{lstlisting}[
language=python,
caption={Generated block  (simplified)},
label=lst:generated-block
]
class TransformerBlock(nn.Module):
    def __init__(self, config):
        super().__init__()
        self.attn = FlexGenAttention(config)   # from attention_class
        self.mlp = FlexGenMLP(config)          # from mlp_class
        self.norm = RMSNorm(config.hidden_size)  # from norm_class

    def forward(self, x, kv_cache=None, mask=None):
        if config.rotary_embeddings:
            x = apply_rotary(x)  # from features.rotary_embeddings

        if config.hf_causal_mask:
            mask = build_4d_mask(mask)  # from features.hf_causal_mask

        if config.use_cache:
            kv_cache = update_cache(kv_cache, x)  # from features.use_cache

        out = self.attn(x, kv_cache, mask)
        out = self.mlp(out)
        return out
\end{lstlisting}

\begin{table*}[t]
\centering
\caption{Transport breakdown across network environments for Falcon-7B (2 GPUs, blocks 0:16 and 16:32) at batch size 32, FP16, micro-batching off.}
\label{tab:transport_breakdown_falcon7b_bs32}
\scriptsize
\setlength{\tabcolsep}{3.5pt}
\renewcommand{\arraystretch}{1.08}
\begin{tabular}{lrrrrrrrrrr}
\toprule
\textbf{Env.} & \textbf{$T_{GPU \to CPU}$} & \textbf{$T_{CPU \to NIC}$} & \textbf{$T_{NIC \to NIC}$} & \textbf{$T_{NIC \to CPU}$} & \textbf{$T_{CPU \to GPU}$} & \textbf{Inference Lat.} & \textbf{Throughput} & \textbf{Comm. Vol.} & \textbf{$T_{GPU}$ Compute} & \textbf{Server-side Net Lat.} \\
& \textbf{(ms)} & \textbf{(ms)} & \textbf{(ms)} & \textbf{(ms)} & \textbf{(ms)} & \textbf{(ms)} & \textbf{(tok/s)} & \textbf{(KB)} & \textbf{(ms)} & \textbf{(ms)} \\
\midrule
E1 & 1.23 & 0.25 & 8.28 & 0.07 & 0.60 & 267.61 & 119.38 & 284.4 & 35.47 & 8.36 \\
E2 & 1.32 & 0.27 & 36.02 & 0.13 & 0.44 & 287.34 & 111.87 & 284.4 & 36.65 & 36.15 \\
E3 & 1.26 & 0.26 & 63.47 & 0.08 & 0.44 & 291.70 & 107.02 & 284.4 & 35.41 & 63.54 \\
E4 & 1.25 & 0.26 & 98.79 & 0.08 & 0.46 & 327.04 & 97.30 & 284.4 & 36.15 & 98.87 \\
E5 & 1.23 & 0.26 & 220.90 & 0.08 & 0.48 & 477.22 & 66.23 & 284.4 & 35.21 & 220.98 \\
\bottomrule
\end{tabular}
\end{table*}

\begin{table*}[t]
\centering
\caption{Transport breakdown across network environments for Mixtral-8$\times$7B (4 GPUs, blocks 0:8, 8:16, 16:24, 24:32) at batch size 32, FP16, micro-batching off. Timing metrics are from the first server (S1$\to$S2 push).}
\label{tab:transport_breakdown_mixtral_bs32}
\scriptsize
\setlength{\tabcolsep}{3.5pt}
\renewcommand{\arraystretch}{1.08}
\begin{tabular}{lrrrrrrrrrr}
\toprule
\textbf{Env.} & \textbf{$T_{GPU \to CPU}$} & \textbf{$T_{CPU \to NIC}$} & \textbf{$T_{NIC \to NIC}$} & \textbf{$T_{NIC \to CPU}$} & \textbf{$T_{CPU \to GPU}$} & \textbf{Inference Lat.} & \textbf{Throughput} & \textbf{Comm. Vol.} & \textbf{$T_{GPU}$ Compute} & \textbf{Server-side Net Lat.} \\
& \textbf{(ms)} & \textbf{(ms)} & \textbf{(ms)} & \textbf{(ms)} & \textbf{(ms)} & \textbf{(ms)} & \textbf{(tok/s)} & \textbf{(KB)} & \textbf{(ms)} & \textbf{(ms)} \\
\midrule
E1 & 1.25 & 0.25 & 4.01 & 0.07 & 0.43 & 349.89 & 92.98 & 256.4 & 38.28 & 4.08 \\
E2 & 1.24 & 0.25 & 33.31 & 0.08 & 0.44 & 406.76 & 77.93 & 256.4 & 38.12 & 33.39 \\
E3 & 1.25 & 0.25 & 62.42 & 0.08 & 0.45 & 502.54 & 63.28 & 256.4 & 38.54 & 62.49 \\
E4 & 1.26 & 0.25 & 96.60 & 0.08 & 0.45 & 605.42 & 52.75 & 256.4 & 38.76 & 96.68 \\
E5 & 1.32 & 0.25 & 209.70 & 0.11 & 0.46 & 971.91 & 32.70 & 256.4 & 39.55 & 209.81 \\
\bottomrule
\end{tabular}
\end{table*}

\begin{table*}[t]
\centering
\caption{Transport breakdown across network environments for Falcon-40B (3 GPUs, blocks 0:20, 20:40, 40:60) at batch size 32, FP16, micro-batching off. Timing metrics are from the first server (S1$\to$S2 push). Mean of 5 runs after warmup.}
\label{tab:transport_breakdown_falcon40b_bs32}
\scriptsize
\setlength{\tabcolsep}{3.5pt}
\renewcommand{\arraystretch}{1.08}
\begin{tabular}{lrrrrrrr}
\toprule
\textbf{Env.} & \textbf{$T_{GPU \to CPU}$} & \textbf{$T_{CPU \to NIC}$} & \textbf{$T_{NIC \to NIC}$} & \textbf{Inference Lat.} & \textbf{Throughput} & \textbf{Comm. Vol.} & \textbf{$T_{GPU}$ Compute} \\
& \textbf{(ms)} & \textbf{(ms)} & \textbf{(ms)} & \textbf{(ms)} & \textbf{(tok/s)} & \textbf{(KB)} & \textbf{(ms)} \\
\midrule
E1 & 1.31 & 0.39 & 5.64 & 506.2 & 63.18 & 512.4 & 61.80 \\
E2 & 1.35 & 0.37 & 38.14 & 578.1 & 55.32 & 512.4 & 63.70 \\
E3 & 1.40 & 0.39 & 72.30 & 625.4 & 51.15 & 512.4 & 62.74 \\
E4 & 1.34 & 0.36 & 114.66 & 718.6 & 44.52 & 512.4 & 61.76 \\
E5 & 1.34 & 0.38 & 316.42 & 1118.0 & 28.62 & 512.4 & 57.00 \\
\bottomrule
\end{tabular}
\end{table*}

\subsection{Micro-batching on Falcon-7B and Mixtral-8$\times$7B}
\label{sec:appendix_microbatch}

We evaluate micro-batching on Falcon-7B and Mixtral-8$\times$7B to examine its sensitivity to pipeline depth. Table~\ref{tab:mb_falcon_mixtral} summarizes the throughput with and without micro-batching across E1--E5. While the overhead in E1 is similar for both models ($\approx$18--19\%), the benefit in E5 increases with deeper pipelines: from +0.9\% for Falcon-7B (2 stages) to +11.4\% for Mixtral-8$\times$7B (4 stages).

\begin{table*}[t]
\centering
\small
\caption{Micro-batching effect on Falcon-7B (2 GPUs) and Mixtral-8$\times$7B (4 GPUs). All runs use batch size 32, FP16, sequence length 128. Throughput in tokens/s.}
\label{tab:mb_falcon_mixtral}
\setlength{\tabcolsep}{3pt}
\renewcommand{\arraystretch}{1.0}
\begin{tabular}{l rr r rr r}
\toprule
& \multicolumn{3}{c}{\textbf{Falcon-7B (2 GPUs)}} & \multicolumn{3}{c}{\textbf{Mixtral-8$\times$7B (4 GPUs)}} \\
\cmidrule(lr){2-4} \cmidrule(lr){5-7}
\textbf{Env.} & \textbf{MB off} & \textbf{MB on} & \textbf{$\Delta$} & \textbf{MB off} & \textbf{MB on} & \textbf{$\Delta$} \\
\midrule
E1 & 119.38 & 97.37 & $-$18.4\% & 92.98 & 75.55 & $-$18.7\% \\
E2 & 111.87 & 91.33 & $-$18.4\% & 77.93 & 66.35 & $-$14.9\% \\
E3 & 107.02 & 86.75 & $-$18.9\% & 63.28 & 58.09 & $-$8.2\% \\
E4 & 97.30 & 79.11 & $-$18.7\% & 52.75 & 49.55 & $-$6.1\% \\
E5 & 66.23 & 66.80 & +0.9\% & 32.70 & 36.42 & +11.4\% \\
\bottomrule
\end{tabular}
\end{table*}